\documentclass[amsmath,amssymb,  aps,  twocolumn, superscriptaddress, floatfix, prl]{revtex4-2}
\usepackage{gensymb}
\usepackage{graphicx}
\usepackage{dcolumn}
\usepackage{bm}
\usepackage{xcolor}
\usepackage{epsfig}
\usepackage[dvipsnames]{xcolor}
\usepackage{soul,enumitem,ulem}
\usepackage{hyperref}
\usepackage[caption=false]{subfig}
\newcommand{\heading}[1]{\textit{#1}}
\newcommand{\Wi}{\mathrm{Wi}}
\newcommand{\Rey}{\mathrm{Re}}
\newcommand{\El}{\mathrm{El}}
\usepackage{lipsum}

\begin{document}


\title{Experimental Evidence of Elastic Instability in Viscoelastic Kolmogorov Flow}
\author{Christophe Brouzet} 
\altaffiliation{equal contribution}
\affiliation{Université Côte d’Azur, CNRS, Institut de Physique de Nice (INPHYNI),  Nice 06200, France}

\author{Mayank Chopra} 
\altaffiliation{equal contribution}
\affiliation{Department of Chemical Engineering, Indian Institute of Technology Bombay,   Mumbai 400076, India}

\author{Dario Vincenzi} 
\affiliation{Université Côte d’Azur, CNRS, LJAD, Nice 06100, France}

\author{Jason R. Picardo} 
\email{picardo@iitb.ac.in}
\affiliation{Department of Chemical Engineering, Indian Institute of Technology Bombay,   Mumbai 400076, India}

\author{Atul Varshney} 
 \email{atul.varshney@univ-cotedazur.fr}
\affiliation{Université Côte d’Azur, CNRS, Institut de Physique de Nice (INPHYNI),  Nice 06200, France}


\begin{abstract}
Inertialess elastic instabilities can profoundly affect transport in complex-fluid flows, from mixing in microfluidic devices and displacement through porous media to polymer processing. While such instabilities are well understood in flows with curved streamlines, their origin in rectilinear flows like the prototypical Kolmogorov flow remains unresolved. Here, we report the first experimental evidence of the center-mode elastic instability, at vanishing inertia, in a confined quasi-two-dimensional viscoelastic Kolmogorov flow, driven by electromagnetic forcing. The instability is characterized by the emergence of ``arrow-shaped" structures in the velocity field, which fluctuates chaotically as the forcing strength increases. We reproduce these observations in simulations of the Oldroyd-B model and trace the dynamics in the confined flow to the travelling arrowhead structures that emerge from the center-mode instability of unbounded Kolmogorov flow. Our work establishes a generic route to purely elastic instability in rectilinear flows, broadening the conditions under which elastic turbulence and enhanced mixing can arise.
\end{abstract} 
\maketitle
\heading{Introduction---}Viscoelasticity, a defining feature of many biological and industrial fluids, can give rise to instabilities and fluctuating flows even in the absence of inertia~\citep{shaqfeh_purely_1996, steinberg_elastic_2021}. Understanding such elastic instabilities is a prerequisite for
both harnessing them in small-scale mixing devices and suppressing them in rheological measurements. Theory, simulations, and experiments have established a clear and consistent picture of elastic instabilities in flows with curvilinear streamlines, such as that between parallel discs and in serpentine channels~\citep{shaqfeh_purely_1996,McKinley96,groisman_elastic_2000, groisman_elastic_2004,Dutta_et_al,Castillo_et_al, steinberg_elastic_2021}; the stretching of the microstructure (e.g., polymers) along the streamlines generates elastic hoop stresses, whose nonlinear coupling with the flow enhances perturbations.

Our understanding is, however, far from complete in the case of parallel shear flows with rectilinear streamlines~\citep{Dutta_et_al,Castillo_et_al}.
In place of the hoop-stress instability,  various alternative mechanisms have been proposed for the observed destabilization of the flow in straight pipes and channels~\citep{pipeflow_Kellay_2011,jha_steinberg_2021,shnapp_steinberg_2022, li_steinberg_2025_b}. These include nonlinear~\citep{morozov_prl2005,pan_et_al_2013,varshney_elastic_2017, varshney_mixing_2018, varshney_elastic_2019} and linear nonmodal~\citep{nonmodal_Shnapp_2022} mechanisms, as well as two linear modal mechanisms---the center-mode~\citep{khalid_centre-mode_2021, Khalid_PRL2021} and polymer-diffusive \citep{BeneitezPRF2023,Lewy} instabilities. In channel flow, the center-mode instability is characterized by perturbations that propagate with a phase speed close to the maximum base-flow velocity and  are localized near the  centerline, where the base-flow velocity is maximal and the shear rate vanishes \citep{Khalid_PRL2021,kerswell_asymptotics_2024}. It has attracted considerable attention in recent years, notably because it gives rise to travelling coherent structures, known as ``arrowheads'' or ``narwhals''~\citep{Morozov_2022,Buza_Page_Kerswell_2022,Buza_Beneitez_Page_Kerswell_2022}, whose chaotic interactions can sustain elastic turbulence in channel flows \citep{Lellep_2023,morozov2025blessings}. Similar center-mode-driven structures also develop in the presence of inertia \citep{garg_prl,chaudhary_pipe_jfm_2021,Buza_Page_Kerswell_2022,Buza_Beneitez_Page_Kerswell_2022, dubief_first_2022,Khalid_PRL2021,choueiri_experimental_2021}.

Development of the theory of a center-mode-driven transition to elastic turbulence has greatly benefited from studies in the idealized two-dimensional viscoelastic Kolmogorov flow (vKf), which consists of alternating shear layers maintained by a unidirectional sinusoidal forcing. Indeed, well before the center-mode and narwhals were found in channel flow, they had been discovered in the vKf~\citep{BOFFETTA_et_al_JFM_2005,bistagnino,Berti_PRE2008,Berti_PRE2010}, though this connection was made only recently~\citep{kerswell_asymptotics_2024,Lewy_Kerswell_2025}. Detailed simulations of the two-dimensional vKf have  shown that the center-mode instability produces arrowheads~\citep{Lewy_Kerswell_2025,Kerswell2026TW} that exhibit complex nonlinear dynamics, including period doubling~\citep{Thomases25} and bursting routes~\citep{Lewy_Kerswell_2025} to chaos, ultimately driving elastic turbulence~\citep{Planetal2017,garg_2018,garg_2021,Kerswell26building-blocks}.
While this theoretical progress has advanced our understanding of rectilinear elastic instability, it relies almost exclusively on simple constitutive models and experimental verification is lacking.

In this Letter, we provide the first experimental evidence of  the center-mode elastic instability, at vanishing inertia, in the vKf. The instability gives rise to chaotic fluctuations of the kinetic energy and is characterized by ``arrow-shaped" structures in the velocity field. These structures are shown to be markedly distinct from the wavy, modulated-flow structures produced by the inertial instability of the Newtonian flow, which has been reported previously \citep{tithof_bifurcations_2017}. Two-dimensional (2D) wall-bounded simulations using the Oldroyd-B model confirm the emergence of arrow-shaped structures in the velocity field and additionally reveal distorted arrowhead-like features in the polymer-extension field. By simulating domains of varying size, we show how travelling arrowheads or narwhals---found previously in unbounded domains---manifest in a confined system such as our experiment. 

\begin{figure}[t]
    \centering
        \includegraphics[width=1\columnwidth]{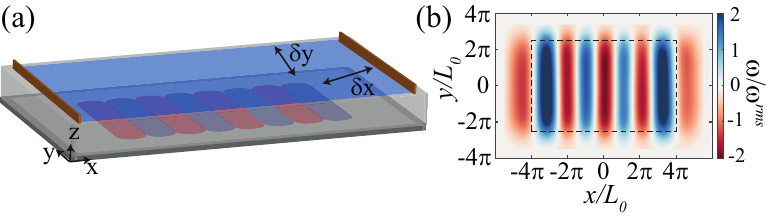}
    \caption{(color online). (a) A schematic view of the two-dimensional Kolmogorov flow setup (not to scale). A magnetic array is placed beneath the flow cell, consisting of  alternating polarity  permanent magnets—indicated in blue (north pole) and red (south pole). A direct current  with  density $\mathbf{J} = J\hat{\mathbf{x}}$ is injected through the electrolyte {\it via} two electrodes (orange rectangles). The interaction of the electric current with the longitudinal magnetic field $\mathbf{B} = B\hat{\mathbf{z}}$ generates a Lorentz force $\mathbf{F} = \mathbf{J} \times \mathbf{B}$, which drives the fluid flow along the $y$-direction. A buffer region of $\delta x = 20$ mm and $\delta y = 15$ mm is maintained between the cell boundaries and the edges of the magnet array to allow the flow to decay smoothly towards the boundaries. (b) Snapshot of time-averaged vorticity field normalized by its root-mean-square value, $\omega/\omega^{rms}$, for the base steady flow at $\Rey=0.3$. The dashed rectangle indicates the position of the magnet array.}
    \label{fig1}
\end{figure}

\heading{Experiments---}A quasi-two-dimensional flow is generated by electromagnetic (Lorentz) forcing, arising from the interaction between  a DC current flowing through a shallow electrolyte layer  and  a spatially varying  magnetic field. A magnetic array  consisting  of eight N52-grade Neodymium permanent  magnets (each measuring $50\times10\times5$ mm$^3$) arranged with alternating polarity is positioned centrally  beneath the flow cell of dimensions $120\times80\times8$ mm$^3$, made of transparent acrylic glass (Fig.~\ref{fig1}(a)). The forcing length scale is defined as $L_0=2w/2\pi\approx 3.2$~mm; $w=10$~mm being the magnet width (see End Matter). The flow cell is filled with an electrolyte, and a direct current (using  power supply PLH250), injected through  Cu electrodes placed across the cell, drives the flow primarily within the top fluid layer of thickness $\sim 1$ mm.   

The electrolyte is prepared by dissolving 70 mM sodium sulfate (Na$_2$SO$_4$) in a 70:30 wt$\%$ mixture of glycerol and water (density $\rho_s=1180$ kg/m$^3$). Long-chain, flexible polymer molecules (Polyacrylamide, molecular weight 18 MDa) are added to the electrolyte at a concentration of 80 ppm, which is well below the overlap polymer concentration of 200 ppm \citep{liu_concentration_2009}. The measured viscosity of the Newtonian electrolyte and polymeric fluid at $21\degree$C are   $\eta_s\approx22$ mPa$\cdot$s and $\eta\approx28.6$ mPa$\cdot$s, respectively, and the polymer relaxation time $\lambda\sim 2$~s \citep{liu_concentration_2009} (see End Matter). Thus, the viscosity ratio is given by $\beta=\eta_s/\eta=0.77$.

The fluid surface is seeded with tracer particles (size $20~\mu$m,  Dynoseeds) and illuminated uniformly using a  white LED panel. Particle image velocimetry (PIV) is performed, with the aid of a high-speed camera (Phantom TE2010) fitted with a Nikkor $50$ mm lens, to obtain spatiotemporally-resolved 2D velocity fields $\bm{u}=(u_x, u_y)$ at the free surface of the fluid (see End Matter). The time-averaged vorticity field $\omega=\nabla\times\bm{u}$, measured when the base  flow is stable, is shown in Fig.~\ref{fig1}(b). 

\heading{Numerical simulations---}We perform simulations of a viscoelastic fluid in a 2D domain of size $8 \pi L_0 \times 8 \pi L_0 $ with walls at $y = \pm 4\pi L_0$ and periodic boundaries at $x = \pm 4\pi L_0$. To emulate the electromagnetic forcing of the experiment, we apply a modified Kolmogorov body force of the form $M(y) \sin(x/L_0) \bm{e}_y$, where the masking function $M$ is unity away from the walls and falls off rapidly to zero near the walls, leaving a buffer length of $\delta y/L_0=1.5 \pi$. The open-source pseudospectral code Dedalus (version 2) \citep{dedalus2020} is used to  numerically solve  the Navier-Stokes equations for the velocity field $\bm{u}$, coupled to the Oldroyd-B model for the polymer conformation tensor $\bm{C}$, using solvent and polymer properties matched to the experiments. The simulations are based on a Chebyshev-Fourier spectral discretization of space, on a $256^2$ grid, and a fourth-order semi-implicit time-stepper \citep{Wang-time-stepper} (see End Matter).

\begin{figure}[t!]
\centering
\includegraphics[width=1\columnwidth]{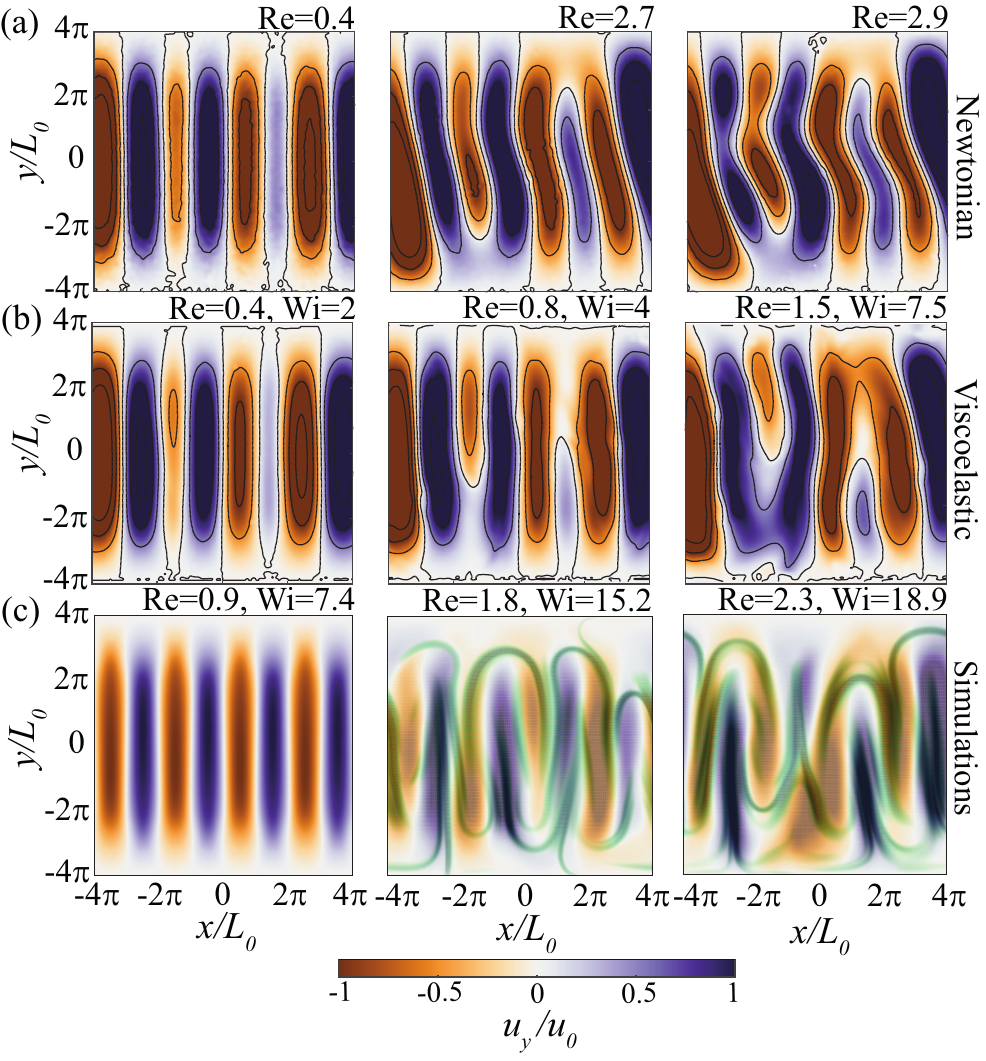}
\caption{(color online). Colormap of the rescaled instantaneous longitudinal velocity $u_y/u_0$ field  obtained from experiments of (a) Newtonian  and (b) viscoelastic Kolmogorov flows. The velocity contours are shown as black lines and the flow fields are displayed between $x/L_0=\pm 4\pi$ (see the dotted region in Fig.~\ref{fig1}(b)). (c) Colormap of $u_y/u_0$ from 2D numerical simulations of the vKf; regions where the log($\operatorname{tr}(\boldsymbol{C})$) exceeds 59\% of its maximum value are shaded in green in the middle and rightmost panels. In all cases, the snapshots from left to right in each row demonstrate the evolution of flow structures with increasing $\Rey$ and $\Wi$,  corresponding to the flow before, shortly after, and well-beyond the instability onset.} 
\label{fig2}
\end{figure}

\heading{Base steady flow---}Under weak driving, the flow closely follows the imposed forcing (Fig.~\ref{fig1}(b)), so that the velocity field in the central portion of the domain is well approximated by $\bm{u} = (0,u_y)$, with $u_y = u_0 \sin(x/L_0)$, corresponding to the  steady base Kolmogorov flow~\citep{Berti_PRE2008}. The speed $u_0$ of the base flow is calculated from the root-mean-square (RMS) of $u_y$ (averaged over the rectangular region in Fig. \ref{fig1}(b)) via $u_0 = \sqrt{2} u_y^{rms}$ and is used to define the Reynolds number $\Rey=u_0 L_0/\nu$, where $\nu = \eta/\rho_s$ is the kinematic viscosity of the polymer solution. We also define the Weissenberg number, $\Wi = \lambda u_0 / L_0$,  as the ratio of the polymer relaxation time $\lambda$ to the characteristic shear timescale $L_0/u_0$. In the experiments, increasing the driving force increases $u_0$, thereby increasing both $\Rey$ and $\Wi$, while leaving  the elasticity number, $\El=\Wi/\Rey=\lambda\nu/ L_0^2 \approx 5$, unchanged.

\heading{Inertial and elastic flow transitions---}In Figs.~\ref{fig2}(a,b), we contrast the flow structures observed for the Newtonian and viscoelastic fluids in the experiments.  In both cases, the base state consists of straight bands of alternating longitudinal flow, as shown in the instantaneous snapshots of $u_y/u_0$ in the first panels of Figs.~\ref{fig2}(a,b). In the Newtonian case (Fig.~\ref{fig2}(a)), the base flow which is stable at low-$\Rey$ (first snapshot at $\Rey=0.4$), undergoes a qualitative transition as $\Rey$ is increased: the longitudinal flow bands become wavy or modulated (middle snapshot at $\Rey=2.7$) to an extent that increases with the forcing strength (third snapshot at $\Rey=2.9$). This inertial ``modulated-flow'' instability has been reported previously and studied extensively through experiments and numerical simulations \citep{tithof_bifurcations_2017}. 

Turning to the viscoelastic case (Fig.~\ref{fig2}(b)), we observe a qualitatively distinct transition  at $\Rey < 1$, i.e., in a regime where the Newtonian flow remains stable. This elastic instability, triggered by increasing $\Wi$, develops through the deformation and closure of same-sign flow bands, leading to the formation of ``arrow-shaped'' structures (middle snapshot at $\Wi=4$ and $\Rey=0.8$).
 
 Notably, these structures repeat  in the $x$-direction, with a spatial period of about $4\pi$, i.e., twice the forcing scale. Further, they are antisymmetric (pointing in opposite directions) about the $x=0$ plane. This is also seen in the instantaneous profile of the transverse velocity $u_x/u_0$ along the centerline $y/L_0=0$ (see End Matter). As $\Wi$ increases, these arrow-shaped structures become more pronounced (third snapshot at $\Wi=7.5$ and $\Rey=1.5$). 

The simulations exhibit similar arrow-shaped structures in the velocity field at the onset of elastic instability, as shown in Fig.~\ref{fig2}(c). Moreover, the field of the polymer's squared extension  $\operatorname{tr}(\boldsymbol{C})$ is concentrated in arrowhead-like structures, shown by the green overlay   highlighting regions of large extension, which appear to correlate with the arrow-shaped structures in the velocity field. These arrowhead-like filaments in the $\operatorname{tr}(\boldsymbol{C})$ field are reminiscent of the travelling-wave patterns known as arrowheads or narwhals, which have been theoretically predicted to arise via the center-mode instability in  rectilinear viscoelastic flows, i.e., including both Kolmogorov~\citep{BOFFETTA_et_al_JFM_2005,bistagnino,Berti_PRE2008,Berti_PRE2010,kerswell_asymptotics_2024,Lewy_Kerswell_2025} and channel flows~\citep{khalid_centre-mode_2021,Khalid_PRL2021,Morozov_2022,Buza_Page_Kerswell_2022,Buza_Beneitez_Page_Kerswell_2022}. We show later how arrowheads can manifest despite the presence of boundaries in the flow direction, albeit in a distorted form as in Fig.~\ref{fig2}(c).

\heading{Bifurcation diagrams---}Next, we quantitatively characterize the observed
inertial and elastic flow transitions (Fig.~\ref{fig2}). In the base state, wherein the flow bands are straight and parallel, the transverse velocity $u_x$ is negligibly small compared to the longitudinal velocity $u_y$, in the central region of the domain (see  End Matter). As the instability sets in, however, the bands deform and the magnitude of $u_x$ increases (see End Matter). We therefore use $u_x^{rms}/u_0$ as an order parameter to characterize the flow transitions; here $u_x^{rms}={\langle u_x^2\rangle}^{1/2}$ with the average taken over time and over a $5\pi\times\pi$ rectangular region centered at $(x/L_0,y/L_0)=(0,0)$ (see  End Matter). 
 
\begin{figure}[t!]
\centering
\includegraphics[width=1\columnwidth]{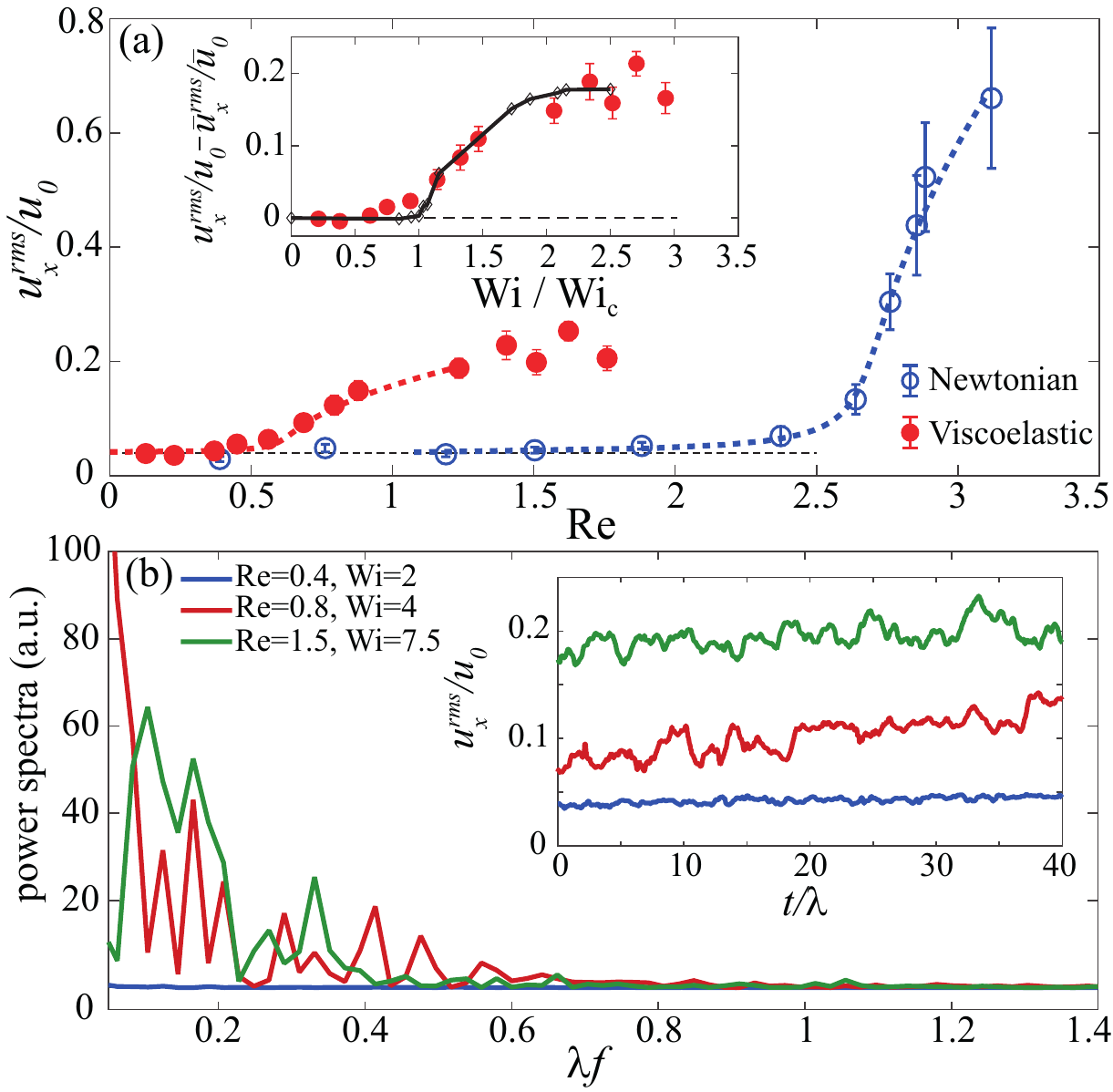}
\caption{(color online). (a) Variation of the order parameter~$u^{rms}_x/u_0$ with $\Rey$, characterizing the growth of elastic (red symbol) and Newtonian (blue symbol) instabilities. The  dashed red and blue lines indicate the  data fitted using the Landau equation (see text),
over a narrow range of $\Rey$ above the onset of the elastic and Newtonian instabilities, respectively. This yields instability  onset values   $\Rey_c\approx0.6$ and 2.7 for viscoelastic and Newtonian fluids, respectively. Inset: The variation of $u^{rms}_x/u_0$, after subtracting the base flow value $\overline{u}^{rms}_x/\overline{u}_0$, as a function of $\Wi/\Wi_c$ for viscoelastic fluid with $\Wi_c^{exp}\sim3$. The growth behavior of the elastic instability is in excellent  agreement with the simulations  with $\Wi_c^{sim}\sim10$ (black curve). The error bars on $u^{rms}_x/u_0$ are estimated from the standard deviation of $u_x/u_0$. (b) The frequency power spectra of $u^{rms}_x /u_0$  as a function of $\lambda f$, exhibiting broadband, aperiodic fluctuations above $\Wi_c$. Inset: The corresponding time series of $u^{rms}_x /u_0$ for three $\Wi$.}
\label{fig3}
\end{figure}

 The main panel of Fig.~\ref{fig3}(a) presents the variation of $u_x^{rms}/u_0$ with $\Rey$. For the Newtonian fluid (blue curve), $u_x^{rms}/u_0$ remains nearly constant, close to its base-flow value, up to $Re\approx2.4$ and then begins to grow  as the modulated flow structures emerge in the velocity field. This transition has been reported previously and is attributed to an inertia-driven pitchfork bifurcation \citep{tithof_bifurcations_2017}. 

\begin{figure}[t]
\centering
\includegraphics[width=\linewidth]{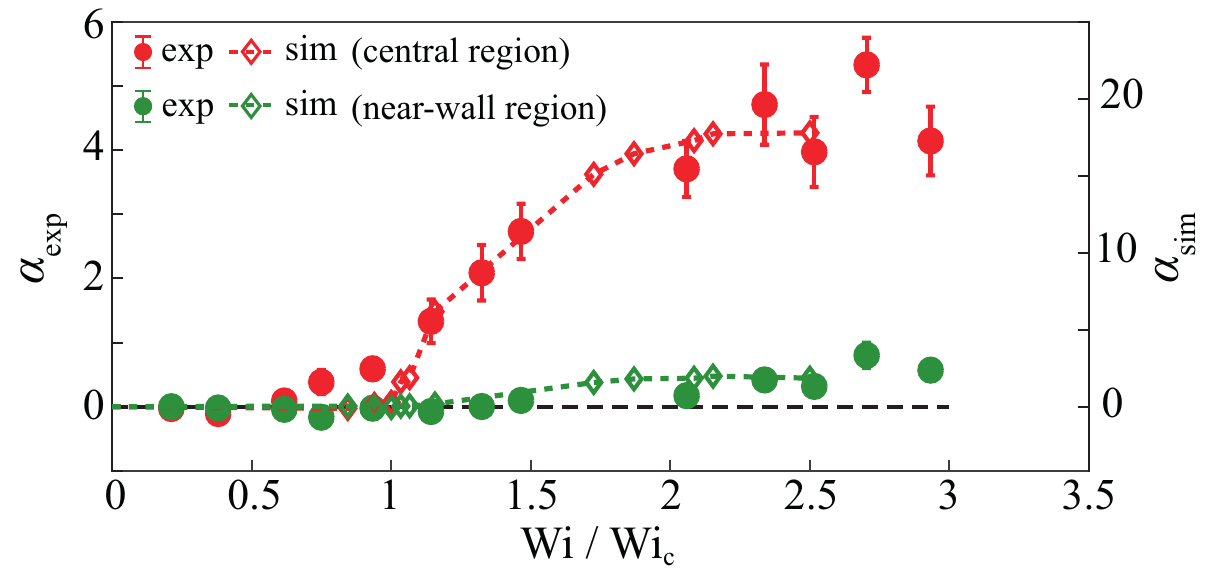}
\caption{(color online). Variation of $\alpha=({u^{rms}_x/u_0})/(\overline{u}^{rms}_x/\overline{u}_0
)-1$ as a function of $\Wi/\Wi_c$ for viscoelastic fluid in the central ($y/L_0=0$;  red)  and the near-wall ($y/L_0=2.5\pi$;  green) regions. Experimental results are shown as solid symbols (left axis), while simulation results are shown as dashed lines (right axis). Note that the smaller experimental values of $\alpha$ arise from the lower spatial resolution of the velocity-field measurements, which results in larger base-state values of $u_x$ than those in the simulations.}
\label{fig8}
\end{figure}
In case of the viscoelastic fluid (red curve; Fig.~\ref{fig3}(a)), the transition occurs at a much lower $\Rey$, below unity, beyond which $u_x^{rms}/u_0$ grows nonlinearly. Importantly, the Newtonian base flow is still stable at these values of $\Rey$. This observation confirms that, in the viscoelastic fluid, the instability and the resulting flow structures and dynamics are driven solely by elastic stress rather than inertia, even well beyond the onset, where $\Rey$ is slightly above unity. Although the growth of $u_x^{rms}/u_0$ is weaker for the elastic transition than for its Newtonian counterpart, the near-onset growth  is well described in both cases by the Landau equation for a continuous, supercritical pitchfork bifurcation: $[(Re/Re_c)-1](u_x^{rms}/u_0)-c(u_x^{rms}/u_0)^3+d=0$, where $c$ and $d$ are constants \citep{varshney_elastic_2017,cross_pattern_1993}. A curve-fit (dashed lines in Fig.~\ref{fig3}(a)) yields critical transition values of $Re_c=2.7\pm0.2$ and $0.6\pm0.1$ for the Newtonian and elastic instabilities, respectively.

The elastic transition is reproduced by our simulations.  In Fig.~\ref{fig3}(a) inset, we plot $u_x^{rms}/u_0$, after subtracting the base flow value $\overline{u}_x^{rms}/\overline{u}_0$ (at $\Wi<\Wi_c$), as a function of $\Wi/\Wi_c$ for the experiments (red symbols) and the simulations (black line). The critical values of $\Wi$ at the onset of the elastic instability are $\Wi_c^{exp}\sim3$ and $\Wi_c^{sim}\sim10$ for the experiments and simulations, respectively.  Although $\Wi_c^{sim}>\Wi_c^{exp}$, the simulations qualitatively capture the instability growth despite the simplifying assumptions of a purely 2D flow of an Oldroyd-B fluid with a single value of $\lambda$ (the polydisperse polymer solution used in the experiments naturally exhibits a distribution of relaxation timescales \citep{Berman1977,BRANDFELLNER2024,serafini_role_2025}). We further repeat the simulations using the Stokes equations (by setting $\Rey = 0$), which leaves the qualitative features of this transition unchanged (results not shown), further confirming its elastic origin.
\begin{figure*}[t]
\centering
\includegraphics[width=0.6\linewidth]{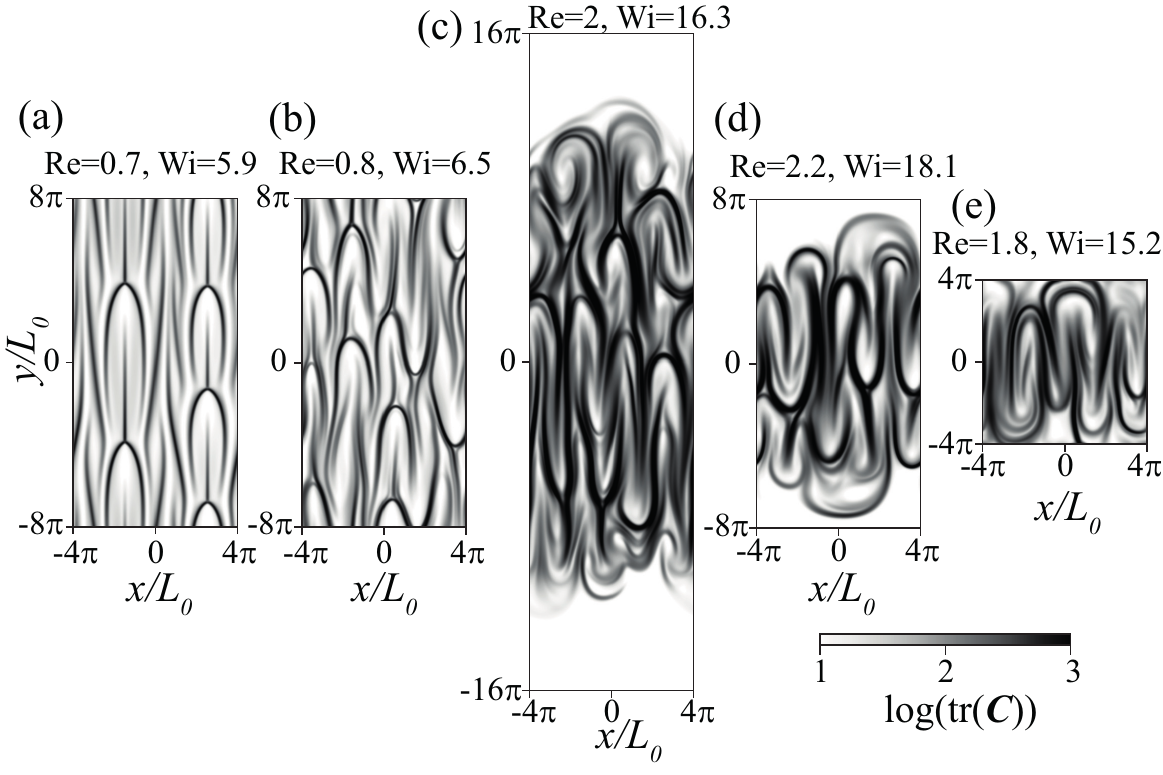}
\caption{Snapshots of the logarithm of the polymer's squared-extension field, log($\operatorname{tr}(\boldsymbol{C})$), in the stationary regime of simulations performed for different domain sizes and boundary conditions. Periodic boundaries in all directions are used for panels (a) and~(b), while walls are located at the longitudinal ends of the domain in (c) to (e).}
\label{fig4}
\end{figure*}

\heading{Chaotic dynamics---}Figure \ref{fig3}(b) examines the temporal dynamics of the viscoelastic flow at three values of $\Wi$ (corresponding to the three panels in Fig.~\ref{fig2}(b)). The temporal fluctuations of $u_x^{rms}/u_0$, presented in the inset, are negligible for the stable base state (blue curve for $\Wi = 2$); in fact, these fluctuations are attributable to noise in the PIV velocity measurements, as evidenced by the associated flat power spectrum (Fig. \ref{fig3}(b) main panel). After the instability, the temporal fluctuations are much stronger and the spectra exhibit multiple peaks (respective red and green curves for $\Wi = 4$ and $7.5$). The aperiodic  temporal fluctuations, evidenced by non-harmonic peaks in the power spectra, indicate that the flow becomes chaotic owing to the dynamics of the emergent arrow-shaped  structures (Fig. \ref{fig3}(b)). Our simulations also predict chaotic fluctuations in $u_x^{rms}/u_0$, as well as in the averaged $\operatorname{tr}(\boldsymbol{C})$, for $\Wi$ above the elastic instability onset (see End Matter).

\heading{Central and near-wall regions---}Could the confining walls be responsible for the observed dynamics? The presence of curvilinear streamlines in the recirculation zones near the confining walls at $y = \pm 4 \pi$  raises the possibility of hoop-stress driven fluctuations at the wall, which are then advected to the central region. However, this scenario is rendered unlikely by the gap $\delta y/L_0$ between the magnet array and the wall (see Fig.~\ref{fig1}), which allows the velocity to decay smoothly and thus keeps the shear stress  much lower near the walls. In fact, on comparing the relative variation of $u_x^{rms}/u_0$ with respect to its base state, defined as $\alpha=({u^{rms}_x/u_0})/(\overline{u}^{rms}_x/\overline{u}_0)-1$, in the central ($y/L_0=0$) and near-wall ($y/L_0=2.5\pi$) regions, we find that only the former grows above the instability onset, whereas the latter remains nearly constant and close to zero, and increases only weakly when the flow becomes chaotic (see  Fig. \ref{fig8}). The simulations also reproduce this distinct behavior of $\alpha$ in the central and near-wall regions.
Therefore, the dynamics must be driven by the center-mode mechanism \citep{kerswell_asymptotics_2024}, which operates in the central region where the base flow is akin to the vKf, with rectilinear streamlines and a velocity maximum at which the shear rate vanishes. 

\heading{Distorted narwhals---}Finally, we investigate how  wall confinement affects the flow structures and the persistence of travelling arrowheads or narwhals under experimentally relevant confinement.
We perform simulations on flow domains with different boundary conditions and sizes, ranging from established unbounded domains to experimentally relevant confined domains. We first consider a long biperiodic domain and a $\Wi$ value just beyond the instability of the base flow. Figure~\ref{fig4}(a) shows a snapshot of the corresponding  $\operatorname{tr}(\boldsymbol{C})$, which clearly exhibits multiple travelling narwhal structures, each characterized by a visible long `tusk'. 

At larger $\Wi$, the interactions between them distort their forms while producing aperiodic fluctuations (Fig.~\ref{fig4}(b)), in agreement with previous work on the 2D unbounded, periodic vKf \citep{Lewy_Kerswell_2025}. Next, we consider a flow domain whose  top and bottom walls are twice as far apart, so that the central region has dimensions comparable to those of the biperiodic domain. When  $\Wi$ is large enough for fluctuating dynamics, we find structures in $\operatorname{tr}(\boldsymbol{C})$ (Fig.~\ref{fig4}(c)) that are remarkably similar to the distorted narwhals in the biperiodic domain (Fig.~\ref{fig4}(b)). The narwhals appear in the central region, meander  across the domain, and then fade away as they approach the walls (see Movie~1 in \citep{supp}). Finally, we shrink the length of the domain to match the aspect ratio of the experiment, while adjusting $\Wi$ such that it remains well above the elastic instability onset. Figures~\ref{fig4}(d, e) show that the narwhals are progressively distorted by confinement into short-lived dynamic arrowhead-like structures (see Movie~2 in \citep{supp}),  which also correlate with the contours of the longitudinal velocity $u_y/u_0$ (Fig.~\ref{fig2}(c)). Hence, the region of rectilinear shear, although limited in extent, drives the dynamics through structures that, given sufficient space, would expand into arrowheads.

\heading{Conclusion---}We experimentally identified a purely elastic instability, at vanishing inertia, in a quasi-two-dimensional viscoelastic Kolmogorov flow. At the instability onset, the longitudinal velocity field develops distinctive arrow-shaped structures and subsequently at higher $\Wi$ the flow fluctuates chaotically. The instability dynamics is governed by the center-mode mechanism operating in the central region, where the base flow closely resembles the vKf. Our wall-bounded viscoelastic simulations reproduced these observations and also revealed arrowhead-like structures in the polymer extension field. Furthermore, these structures are shown to be distorted forms of the arrowheads or narwhals that have been predicted to arise from the  center-mode instability in unbounded Kolmogorov and channel flows. This close correspondence  demonstrates that the experimentally observed elastic instability and associated flow structures originate from the center-mode instability.

\heading{Acknowledgments---}We thank Sandra Lerouge and Shailendra Kumar Yadav for helpful discussions, and Sandip Wadhai and Venkatesh Jha for performing initial experiments. Funding is gratefully acknowledged from the   F{\'e}d{\'e}ration Doeblin (A.V. and D.V.) and the Indo-French Centre for the Promotion of Advanced Research (IFCPAR/CEFIPRA) (J.R.P. and D.V., project No. 6704-1). Simulations were performed on the workstation \textit{Faramir}, procured through the IIT Bombay grant RD/0519-IRCCSH0-021. Computing support from the National PARAM Supercomputing Facility \textit{PARAM SIDDHI-AI} at CDAC, Pune is also gratefully acknowledged. A.V. acknowledges  support from the ANR-CPJ (project No. ANR-24-CPJ1-0027-01) and  has greatly benefited from a Visiting Assistant Professor position at  IIT Bombay. 

%

\onecolumngrid
\appendix
\section{\large End Matter}
\twocolumngrid

\begin{figure}[ht]
\centering
\includegraphics[width=1\linewidth]{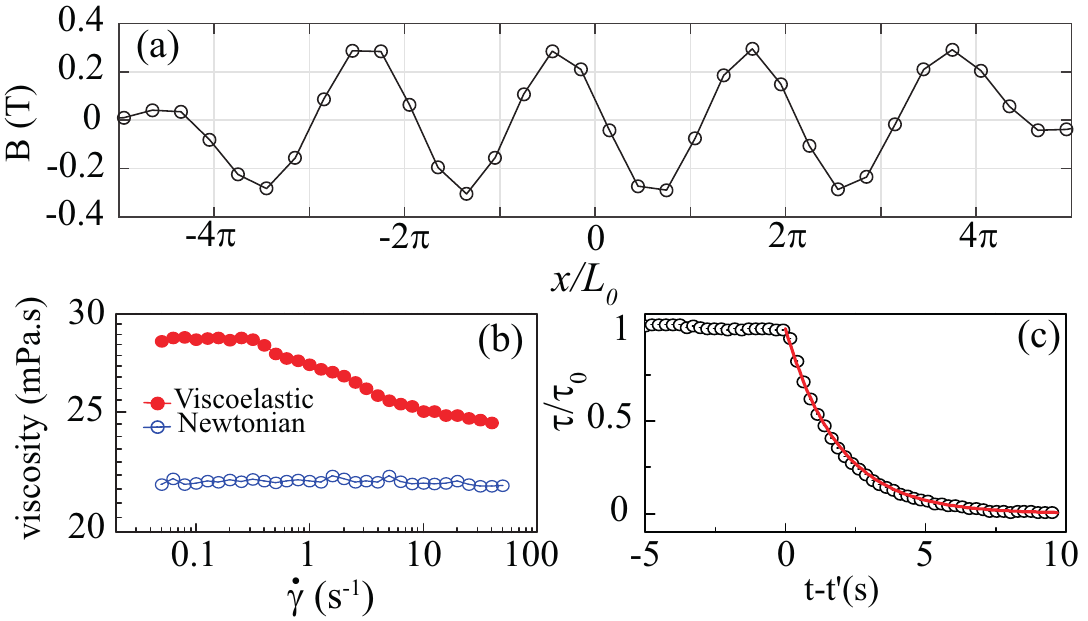}
\caption{(a) Spatial variation of the $z$-component of the magnetic field $B$ along the $x$-direction.  (b) Measured shear viscosity of viscoelastic  and Newtonian  fluids, used in the experiments, as a function of shear-rate $\dot{\gamma}$. (c) Normalized shear stress $\tau/\tau_0$ demonstrating the exponential relaxation of the polymers; $\tau_0$ denotes the shear stress at $\dot{\gamma}=0.5$ s$^{-1}$, which is held constant for $t'=15$ s.  
The red line is an exponential fit to the data of the form $\tau/\tau_0=e^{-t/\lambda}$, yielding $\lambda=2\pm0.2$ s.   } 
\label{fig5}
\end{figure}

 \heading{Forcing scale, rheological characterization and PIV---}The  magnetic field, $B$, measured in air  using a Gauss meter (F.W. Bell model 4048) shows a periodic variation along $x$-direction  (Fig. \ref{fig5}(a)). The magnetic field is measured at $y/L_0 = 0$ and at the bottom surface of the flow cell, i.e. approximately 1 mm above the magnet array.
   
 The  viscosity of the Newtonian electrolyte and the polymer solution are measured  using HR20 rheometer (TA instruments) at $21\degree$C (Fig. \ref{fig5}(b)).  A  dilute addition of polymers increases the solution viscosity~${\eta}$ by approximately $30\%$ and induces shear-thinning behavior at a shear-rate of $\dot{\gamma}\approx0.5$ s$^{-1}$ (Fig. \ref{fig5}(b)).  The longest polymer relaxation time~$\lambda$ is determined using the stress-relaxation method, in which a constant $\dot{\gamma}$ is applied for $t'=15$ s, using a cone-plate geometry in the rheometer,  followed by a sudden cessation of shear while monitoring the decay of the shear stress \citep{liu_concentration_2009}. We obtain $\lambda\sim 2$ s of the polymer solution at $21\degree$C (Fig. \ref{fig5}(c)). 

The PIV vector field computation is carried out using four passes with a final interrogation window of size $16 \times 16~\textrm{pixels}\approx 1.6 \times 1.6$~mm$^2$ and $75\%$ overlap, leading to a resolution of $0.4$~mm in both directions. Additional post-processing and analysis are performed using  PIVMat Toolbox \citep{Moisy2007} in Matlab.

 \begin{figure}[ht]
\centering
\includegraphics[width=1\linewidth]{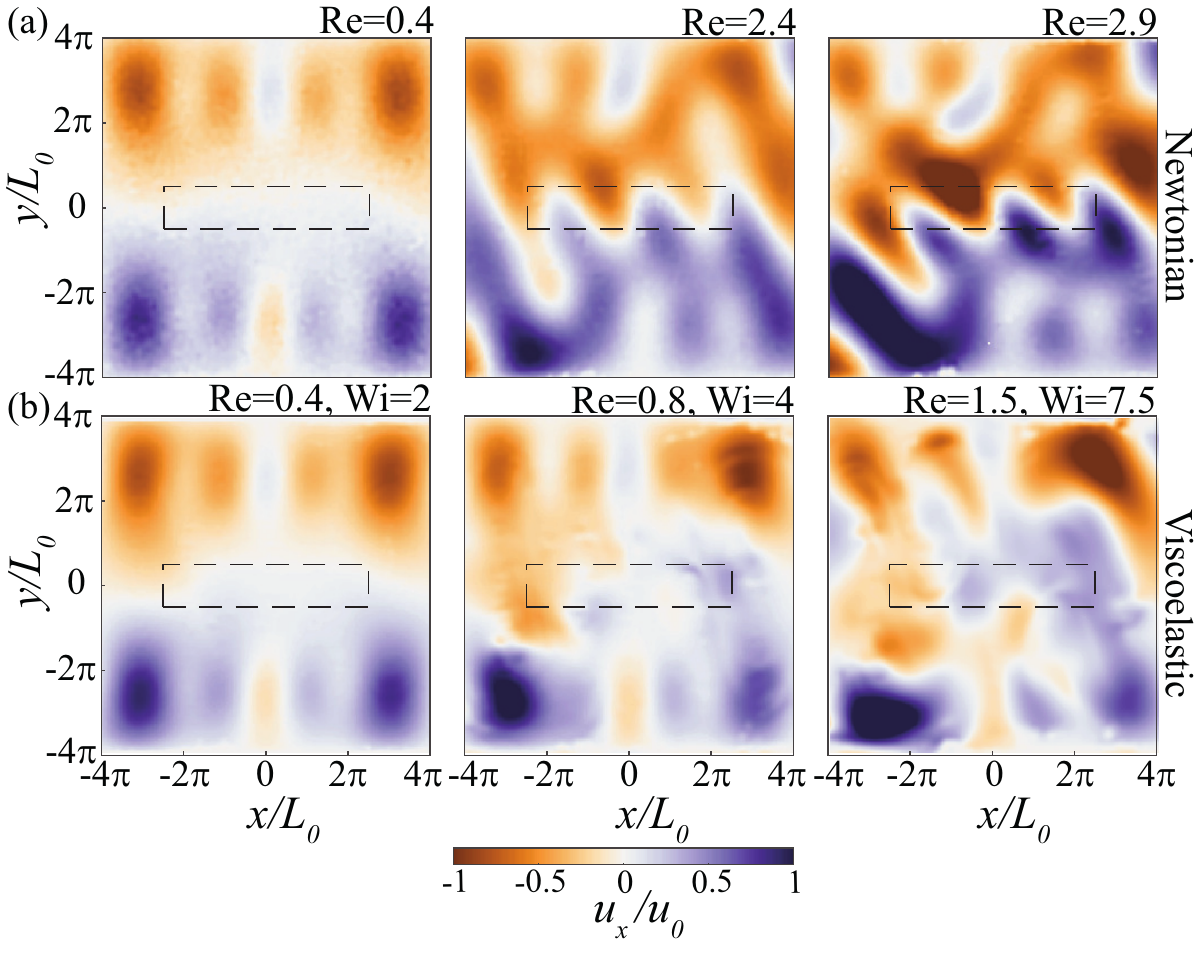}
\caption{Colormap of the rescaled transverse component $u_x/u_0$ of the velocity field  obtained from experiments of (a) Newtonian  and (b) viscoelastic Kolmogorov flows.  In both cases, the snapshots from left to right in each panel demonstrate the evolution of flow structures, from steady to chaotic, with $\Rey$ and $\Wi$ (values are shown atop each snapshot). The dashed rectangle in each snapshot indicates the region over which $u_x$ is spatially averaged.} 
\label{fig6}
\end{figure}
 
 \heading{Transverse velocity field and order parameter---}In Fig.~\ref{fig6}, the  rescaled transverse component $u_x/u_0$ of the velocity field is shown for Newtonian and viscoelastic cases. In the central region (shown by  rectangle) $u_x/u_0$ is nearly zero for the steady base flow, whereas it becomes finite as the instability develops. Thus, the RMS of $u_x/u_0$, averaged over time and over the rectangular region, serves as an order parameter  characterizing both the Newtonian and elastic transitions (Fig. \ref{fig3}(a)).

\begin{figure}[htbp]
\centering
\includegraphics[width=1\linewidth]{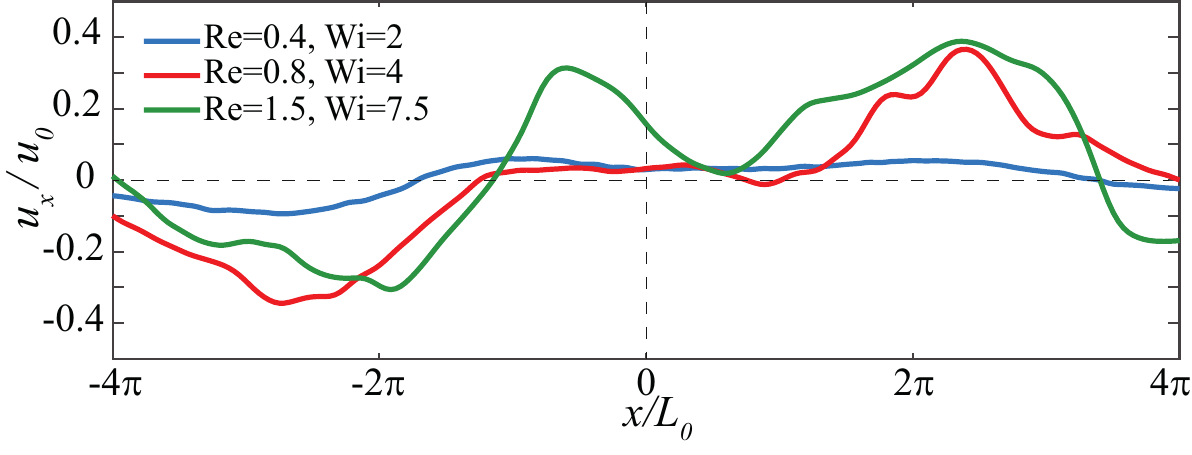}
\caption{Instantaneous transverse velocity $u_x/u_0$ profile measured along the $x$-direction and at $y/L_0=0$ for three $\Rey$ and $\Wi$ values for the viscoelastic fluid. } 
\label{fig7}
\end{figure}

\heading{Transverse velocity profile---}In Fig. \ref{fig7}, we show transverse velocity profiles for the steady base flow and above the elastic instability onset. For the steady flow at $\Wi=2$, the measured profile is nearly flat and remains close to zero, with variations comparable to the experimental noise (blue line). 
Above the elastic instability onset, e.g. at $\Wi=4$ (red line), the profile becomes antisymmetric in the region where the arrow-shaped structures appear, with a characteristic wavelength of about $4\pi$, i.e. double the forcing scale. For high $\Wi$, the arrow-shaped structures exhibit chaotic dynamics, leading to a more complex instantaneous profile ($\Wi=7.5$, green line).

\begin{figure}[t]
\centering
\includegraphics[width=0.6\linewidth]{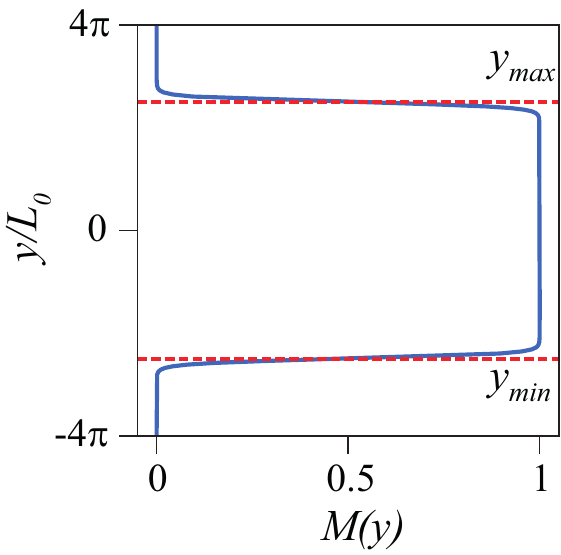}
\caption{Masking function $M(y)$ as a function of the normalized longitudinal coordinate $y/L_0$. The red dashed lines indicate the longitudinal limits, $y_{min}$ and $y_{max}$, of the region over which the Kolmogorov forcing operates. }
\label{fig10}
\end{figure}
 
\heading{Simulation details---}The governing equations for the velocity $\bm{u}$, pressure $p$, polymeric stress $\bm{T}_p$, and conformation tensor $\bm{C}$ are
\begin{gather*}
    \nabla \cdot \bm{u} = 0, \\
        \frac{D \bm{u}}{D t} = -\nabla p + \nu_s \nabla^{2}\bm{u} + \nu_p \nabla \cdot \bm{T}_p + f_0 M(y)\sin{x}\,\bm{e}_y,\\
        \frac{D \bm{C}}{D t} = \bm{C} \cdot \nabla \bm{u} + (\nabla \bm{u})^{T} \cdot \bm{C} + \kappa \nabla^2\bm{C} - (\bm{C} - \bm{I})/\lambda,\\
        \bm{T}_p = (\bm{C} - \bm{I})/\lambda,
\end{gather*}
where $D/Dt$ is the substantial derivative and $\bm I$ is the two-dimensional identity tensor. For Stokes simulations, one must set $D\bm{u}/Dt \rightarrow 0$ in the momentum equation. 

All parameters are fixed except for the forcing amplitude $f_0$ which is varied to increase $\Wi$ (and $\Rey$). The ratio of the kinematic viscosity of the solvent $\nu_s$ to that of the polymer $\nu_p$ is matched to the experiment, $\beta=\eta_s/\eta \approx {\nu_s/(\nu_p+\nu_s)=0.77}$.

Further, $\lambda$ is such that the elasticity number $\El =  8$, which is comparable to the experimental value of 5. Since the diffusivity $\kappa$  of long polymers is typically very small, the inverse P{\'e}clet number $\mathrm{Pe}^{-1}$ takes values in the range of $10^{-5}$ to $10^{-12}$. 
To avoid having to use a very high spatiotemporal resolution, we choose $\kappa$ such that $\mathrm{Pe}^{-1}\approx 10^{-3}$. We have checked and found that our results, especially regarding the elastic instability and the arrowhead-like features in the $\operatorname{tr}(\bm{C})$ field, persist on varying $\kappa$. In fact, we obtain qualitatively similar results (not shown) on setting $\kappa$ to zero and using a finite-difference method with a nondiffusive high-order advection scheme~\citep{Yerasi_Picardo_Gupta_Vincenzi_2024}.

 We perform simulations that are either biperiodic or periodic in the lateral direction with walls at the longitudinal  ends of the domain. For the latter, we set the velocity field and the normal derivative of the conformation tensor to zero at the walls. 
 
 The masking function $M(y)$ ensures that the Kolmogorov forcing decays near the walls in a manner that mimics the experimental configuration, wherein a gap is maintained between the ends of the magnets and the walls. The function $M(y)$, shown in Fig. \ref{fig10}, is constructed from $\tanh$ functions:
    \begin{equation}
        M(y) = \frac{1}{4}\left[1 + \tanh\!\left(\frac{y - y_{\min}}{\delta}\right)\right]  \left[1 + \tanh\!\left(\frac{y_{\max} - y}{\delta}\right)\right] \label{eq:mask}
    \end{equation}
Here, $y_{\min}$ and $y_{\max}$ define the longitudinal limits of the region over which the Kolmogorov forcing is applied, and their values are chosen to match the experimental configuration:
    $y_{\min} = - 2.5\pi L_0$ and $y_{\max} = 2.5\pi L_0$.
The parameter $\delta$ in equation \eqref{eq:mask} smooths the transition between the forced and unforced regions and is chosen as $\delta = 0.094\pi L_0$. We have performed simulations with different gap widths by varying either ($4\pi L_0 - y_{max}$) or ($y_{min}+4\pi L_0$) in the masking function $M(y)$ and found that its precise form does not affect our main result: an elastic instability that gives rise to arrow-shaped structures and fluctuating dynamics.
 \begin{figure}[t]
\centering
\includegraphics[width=1\linewidth]{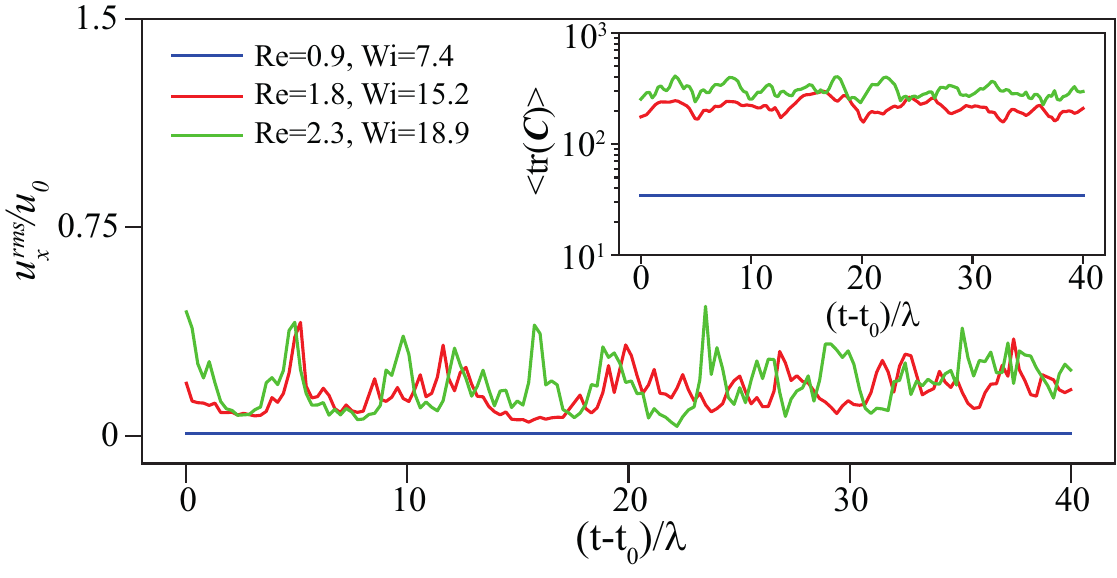}
\caption{Time series of $u_x^{rms}/u_0$, starting from a time $t_0$ in the stationary regime of the simulations, for three values of $\Wi$ (see the legend). Inset: Time series of the spatial-average of $\operatorname{tr}(\boldsymbol{C})$, the squared-extension of the polymer.} 
\label{fig9}
\end{figure}

\heading{Chaotic dynamics in the simulations---}Figure~\ref{fig9} illustrates the temporally fluctuating nature of the flow that arises after the elastic instability onset in the simulations. The main panel shows the time series of $u_x^{rms}/u_0$, starting from a time $t_0$ in the stationary regime, for three values of $\Wi$. While the base flow is steady for $\Wi$ below the transition (blue curve at $\Wi = 7.4$), it exhibits chaotic fluctuations above the transition (respective red and green curves at $\Wi = 15.2$  and $18.9$). Moreover, the time series of spatially-averaged $\operatorname{tr}(\boldsymbol{C})$  also display chaotic fluctuations above  the instability onset, as shown in  the inset of Fig.~\ref{fig9}.
\end{document}